\documentclass[aps,prb,preprint,groupaddress]{revtex4}

\usepackage{graphicx}%

\begin{document}

\title{Classical position probability densities for
spherically symmetric potentials}

\author{Lorenzo J. Curtis}
\email[]{ljc@physics.utoledo.edu}
\author{David G. Ellis}
\email[]{dge@physics.utoledo.edu}

\affiliation{Department of Physics and Astronomy, University of
Toledo, Toledo Ohio 43606}

\date{\today}

\begin{abstract}
A simple position probability density formulation is presented for
the motion of a particle in a spherically symmetric potential. The
approach provides an alternative to Newtonian methods for
presentation in an elementary course, and requires only elementary
algebra and one tabulated integral. The method is applied to
compute the distributions for the Kepler-Coulomb and isotropic
harmonic oscillator potentials. Formulas are also deduced for the
average values for powers of the radial coordinate, and applied to
describe perturbations to these systems. The classical results are
also compared with quantum mechanical calculations using the
Einstein-Brillouin-Keller semiclassical quantization.
\end{abstract}

\maketitle

\section{Introduction}

A significant distinction exists between the conceptual framework
presented in traditional introductory physics courses and that
used in the advanced physics courses that follow them
\cite{Cambridge}. Introductory physics courses utilize historical
Newtonian concepts involving forces and accelerations, but these
concepts never enter in more advanced formulations. The
introductory approach is often characterized as ``classical''
whereas that of the more advanced is described as ``quantum
mechanical.'' However, the primary difference between the two
approaches arises not because of quantization, but instead from a
nonessential heuristic tendency to describe macroscopic systems by
instantaneous values for position, speed, and acceleration, and
microscopic systems by time-averaged position probability
densities.

The reasons for this are clear, since a macroscopic trajectory is
disturbed only slightly when successively interrogated with
visible light, whereas a microscopic system may be destroyed by
interrogation with a single short-wavelength photon.  Thus the
description of the microscopic system requires the superposition
of many similarly interrogated systems. Unfortunately, this
dichotomy produces a serious disconnect between physics as it is
taught to non-major students in service courses and physics as it
is practiced.  Despite efforts to inject modern topics into a
Newtonian presentation, this discontinuity further widens the gap
between physics and society.

In a recent essay, Wilczek \cite{Wilczek} has described the force
concept as an insubstantial ``culture'' that provides a common
language, but not an algorithm for constructing the mechanics of
the world. Similarly, Taylor \cite{Taylor} has suggested an
alternative approach that uses the least action principle in place
of Newtonian forces.  Both essays provide persuasive historical
quotes from respected authorities who have urged that the force
approach to the teaching of elementary physics be replaced.
Unfortunately, the Newtonian model offers practical advantages,
particularly in the testing and evaluation of student performance,
and is thus very firmly entrenched.

It is sometimes argued that initial use of the Newtonian approach
is necessary, because a quantum mechanical formulation would be
too demanding mathematically. However, the problems attacked in
elementary textbooks tend to be simpler than those treated in
quantum mechanical textbooks. If one examines problems of similar
complexity, a Newtonian formulation is often much more complex
mathematically than the corresponding quantum mechanical solution.
For example, elementary textbooks describe the two-dimensional
Kepler orbit problem, but it is invariably restricted to the
special case of a circular orbit (or, in the flat earth
approximation, to a parabolic trajectory). When the classical
problem is formulated in terms of position probability densities,
three-dimensional elliptic orbits are automatically included.
Moreover, deviations from a pure inverse square law can be
included as perturbations \cite{Martin}, all in a purely classical
framework. It is also possible to add semiclassical quantization
directly to the classical solution when desired.

A formulation is presented here in which the periodic
three-dimensional motion of a particle in a central potential is
treated in terms of classical position probability densities. The
method is applied to the problems most frequently encountered in
an introductory quantum mechanics course, namely the
Kepler-Coulomb and isotropic harmonic oscillator potentials. While
these two potentials lead to solutions that possess certain
symmetries, they also have interesting differences.  For example,
the Kepler-Coulomb exemplifies an interaction that decreases with
increasing separation, whereas the isotropic harmonic oscillator
exemplifies an interaction that increases with increasing
separation.

In this presentation the position probability densities are
evaluated, closed form expressions for the average values for
powers of the radial coordinate are obtained, calculations are
made for sample perturbations of the systems, and the connection
to the EBK semiclassical quantization is prescribed.

\section{Position Probability Densities for Central Potentials}

Consider a particle of mass $m$ moving in a central potential $V$($r$%
) described by the standard spherical polar coordinates $r$,
$\vartheta ,\varphi $.  For periodic motion with period $T$, the
dwell time, or position probability density, is given by
\begin{equation}
\label{probr}P(r)dr=\frac{dt}T=\frac 1T\frac{dr}{dr/dt}= \frac
mT\frac{dr}{p_r}
\end{equation}
where $p_r$ is the radial component of the momentum of the
particle, which can be described using conservation of energy as
\begin{equation}
\label{energy}E=\frac{p_r^2}{2m}+\frac{L^2}{2mr^2}+V(r)~.
\end{equation}
Here $L$ is the angular momentum. For a prescribed potential, the
radial momentum can be obtained as
\begin{equation}
\label{momentum}p_r=\sqrt{2m}\sqrt{E-V(r)-L^2/2mr^2}~.
\end{equation}
With periodic orbital motion, the radial coordinate will undergo
librations between turning points that are specified by the roots
of the equation
\begin{equation}
\label{roots}Er^2-V(r)r^2-L^2/2m=0~.
\end{equation}
For the potentials considered here there will be two roots to the
equation, denoted as $A_{\pm }$. Since the potential involves only
$r$, the angular momentum will be constant over the orbit. In the
case of the Kepler-Coulomb and isotropic harmonic oscillator
potentials, the orbits are both ellipses, so Kepler's second law
of equal areas swept out in equal times is valid for both. Thus
\begin{equation}
\label{period}\frac 12r^2\frac{d\varphi }{dt}=\frac{\pi ab}T=\frac L{2m}
\end{equation}
where $a$ and $b$ are the semimajor and semiminor axes of the
ellipse, and $\pi ab$ is its cross sectional area. This equation
permits the specification of the period, which provides the
normalization of the distribution. If $N$
denotes the number of librations in a period ($N$=2 for the Kepler-Coulomb, $%
N$=4 for the harmonic oscillator), then the average values of
powers of $r$ are given by
\begin{equation}
\label{rk}\langle r^k\rangle =\frac NT\int_{A_{-}}^{A_{+}}drP(r)r^k
\end{equation}

\subsection{Kepler-Coulomb Potential}

The potential
\begin{equation}
\label{coulomb}V(r)=-k/r
\end{equation}
gives rise to a negative (binding) energy, which we denote as
$E_B=-E$ so as to explicitly display the sign within square roots.
The momentum thus becomes
\begin{equation}
\label{momcou}p_r=\sqrt{2m}\sqrt{E+k/r-L^2/2mr^2}
\end{equation}
with turning points given by the roots of
\begin{equation}
\label{couroots}-E_Br^2+kr-L^2/2m=0
\end{equation}
given by
\begin{equation}
\label{Kepab}A_{\pm }=\frac k{2E_B}\pm \sqrt{\left( \frac k{2E_B}\right) ^2-%
\frac{L^2}{2mE_B}}.
\end{equation}
In this case the coordinate system is centered on one of the foci
of the ellipse, for which the semimajor and semiminor axes are
given by
\begin{eqnarray}
\label{Kepab2}
a&=&k/2E_B
\nonumber \\
b&=&L\ /\sqrt{2mE_B}~.
\end{eqnarray}
An example of such an orbit with $a=1$ unit and $b=a/2$ is shown
in Fig.~1a.

The period can be computed from the definition of $b$ using Eq.
\ref{period} in the form
\begin{equation}
\label{KepPeriod}T=\frac{2m\pi ab}L=\pi a\sqrt{\frac{2m}{E_B}}~.
\end{equation}
Inserting these relationships into Eq.~\ref{rk} (with $N$=2 since
here the periapsis and apoapsis are separated by 180$^{\rm o})$
\begin{equation}
\label{KepPPD}P(r)dr=\frac 1{\pi
a}\frac{rdr}{\sqrt{(A_{+}-r)(r-A_{-})}}~.
\end{equation}
The position probability density corresponding to the orbit in
Fig.~1a is shown in Fig.~2a.


\subsection{Isotropic harmonic oscillator}

The potential
\begin{equation}
\label{SHOpot}V(r)=kr^2/2
\end{equation}
yields the momentum
\begin{equation}
\label{momSHO}p_r=\sqrt{2m}\sqrt{E-kr^2/2-L^2/2mr^2.}
\end{equation}
with turning points specified by the roots of the equation
\begin{equation}
\label{rootsSHO}Er^2-kr^4/2-L^2/2m=0~,
\end{equation}
given by
\begin{equation}
\label{SHOab}A_{\pm }^2=\frac Ek\pm \sqrt{\left( \frac Ek\right) ^2-\frac{L^2%
}{mk}}.
\end{equation}
This orbit is also elliptical, and is comparable to that of the
Kepler-Coulomb system, except for the fact that the coordinate
system is at the center of the ellipse rather than at one of the
foci. Here the turning points are at the semimajor and semiminor
axes
\begin{eqnarray}
\label{SHOab2}
a&=&A_{+}
\nonumber \\
b&=&A_{-} ~,
\end{eqnarray}
and the period corresponds to four of these turning points. An
example of such an orbit, also with $a=1$ unit and $b=a/2$, is
shown in Fig.~1b.

The area of this ellipse is
\begin{equation}
\label{SHOarea}\pi ab=\sqrt{A_{+}^2A_{-}^2}=\frac L{\sqrt{mk} }~.
\end{equation}
Using Eq. \ref{period}, this gives a value for the period
\begin{equation}
\label{SHOperiod}T=\frac{2m\pi ab}L=2\pi \sqrt{\frac mk}~.
\end{equation}
Inserting these relationships into Eq.~\ref{rk} (with $N$=4 since
here the closest approach and furthest recession are along the
semiaxes, and thus separated by 90$^o)$
\begin{equation}
\label{SHOPPD}P(r)dr=\frac 2\pi
\frac{rdr}{\sqrt{(A_{+}^2-r^2)(r^2-A_{-}^2)}}~.
\end{equation}
The position probability density corresponding to the orbit in
Fig.~1b is shown in Fig.~2b.


\section{Expectation values}

Average values of quantities weighted by these distributions can
be obtained by directly integrating these expressions. However,
they can also transformed into the form of the standard integral
\cite{Gradsteyn}
\begin{equation}
\label{gradsh}\frac 1{2\pi }\int_0^{2\varphi }d\varphi
(1+\varepsilon \cos \varphi )^n=(1-\varepsilon ^2)^{n/2}P_n(\frac
1{\sqrt{1-\varepsilon ^2}})
\end{equation}
where P$_n(x)$ is the Legendre polynomial (in an unusual
application where the argument $x>1$). Negative powers can be
handled using the relationship
\begin{equation}
\label{negn} P_{-n}(x)=P_{n-1}(x)~.
\end{equation}

In addition to the radial integral formulation of Eq.~\ref {rk},
the expectation value can alternatively be written as
\begin{equation}
\label{Pphi}
\langle r^k\rangle =\frac 1T\int_0^Tdt\ r^k=\frac 1T\int_0^{2\pi }\frac{%
d\varphi }{d\varphi /dt}\ r^k ~.
\end{equation}
Conservation of angular momentum relates $r$ and $\varphi $
through Eq.~\ref{period}, which can be rewritten
\begin{equation}
\label{phiconv}T\ d\varphi /dt=2\pi ab/r^2 ~.
\end{equation}
Inserting this into Eq.~\ref{Pphi}
\begin{equation}
\label{phiint}\langle r^k\rangle =\frac 1{2\pi ab}\int_0^{2\pi
}d\varphi \ r^{k+2} ~.
\end{equation}
It remains only to choose the equation of the orbit, and to use
Eq.~\ref{gradsh} to evaluate this expectation value.

\subsection{Kepler-Coulomb problem}

Here the coordinate system is centered on one of the foci of the
ellipse, which has the equation
\begin{equation}
\label{KepEllipse}\frac 1r=\frac a{b^2}(1+\varepsilon \cos \varphi )
\end{equation}
where $\varepsilon \equiv \sqrt{1-b^2/a^2\text{ }}\ $is the
eccentricity of the ellipse. Inserting this relationship for $r$
into Eq.~\ref{phiint}
\begin{equation}
\label{KepInt1}\langle r^k\rangle =\frac 1{ab}\left( \frac a{b^2}\right)
^{-k-2}\frac 1{2\pi }\int_0^{2\pi }d\varphi \ (1+\varepsilon \cos \varphi
)^{-k-2}
\end{equation}
which, using Eq.~\ref{gradsh}, becomes
\begin{equation}
\label{Keprk}\langle r^k\rangle =b^k\left( \frac ba\right)
P_{-k-2}\left( \frac ab\right)~.
\end{equation}
A few examples are:
\begin{eqnarray}
\langle r \rangle ~ &=& a\left[ 3-( b/a)^2\right]/2
\nonumber \\
\langle r^{-1}\rangle &=& 1/a
\nonumber \\
\langle r^{-2}\rangle &=& 1/ab
\nonumber \\
\langle r^{-3}\rangle &=& 1/b^3
\nonumber \\
\langle r^{-4}\rangle &=& \langle r\rangle /b^5 ~.
\end{eqnarray}

\subsection{Isotropic harmonic oscillator problem}

In this case the center of the coordinate is at the center of the
ellipse, and has the equation \cite{Sivard}
\begin{equation}
\label{SHOellipse}\frac 1{r^2}=\frac 12\left( \frac 1{a^2}+\frac
1{b^2}\right) -\left( \frac 1{a^2}-\frac 1{b^2}\right) \cos 2\varphi
\end{equation}
which can be rewritten
\begin{equation}
\label{SHOellip2}\frac 1{r^2}=\left( \frac{a^2+b^2}{4a^2b^2}\right) \left[
1+\left( \frac{a^2-b^2}{a^2+b^2}\right) \cos 2\varphi \right] .
\end{equation}
Defining here
\begin{equation}
\varepsilon \equiv \frac{a^2-b^2}{a^2+b^2}~,
\end{equation}
the quantity occurring in Eq. \ref{gradsh} simplifies to
\begin{equation}
\label{SHOecc}\sqrt{1-\varepsilon ^2}=\sqrt{1-\left( \frac{a^2-b^2}{a^2+b^2}%
\right) ^2}=\left( \frac{2ab}{a^2+b^2}\right)~ .
\end{equation}
The expectation value is given by
\begin{equation}
\label{SHOexp1}\langle r^k\rangle =\frac 1{ab}\left( \frac{a^2+b^2}{2a^2b^2}%
\right) ^{-\frac {k+2}2}\frac 1{2\pi }\int_0^{2\pi }d\varphi \
(1+\varepsilon \cos 2\varphi )^{-\frac {k+2}2}
\end{equation}
which integrates to
\begin{equation}
\langle r^k\rangle =(ab)^{k/2}\ P_{-\frac {k+2}2}\left( \frac {a^2+b^2}{2ab}%
\right)
\end{equation}
This result is valid for both odd and even powers. For odd powers,
the Legendre function can be evaluated numerically as a
hypergeometric series, as shown in the Appendix.

A few examples are:
\begin{eqnarray}
\langle r^2\rangle ~&=& (a^2+b^2)/2
\nonumber \\
\langle r^4\rangle ~&=& \left[ 3\left( a^2+b^2\right)
-4a^2b^2\right]/8
\nonumber \\
\langle r^{-2}\rangle &=& 1/ab
\nonumber \\
\langle r^{-4}\rangle &=& (a^2+b^2)/2a^3b^3 ~.
\end{eqnarray}

\section{Perturbation calculations}

One of the strengths of this method is the ease with which
perturbations to the energy of the system can be computed. The
total energy can be deduced from the potential using the virial
theorem
\begin{equation}
E=\left\langle V(r)\right\rangle +\frac 12\left\langle r\frac {dV}{dr}%
\right\rangle
\end{equation}
so a perturbation of the form $\Delta V(r)$ can be computed as
\begin{equation}
E^{\prime }=E+\left\langle \Delta V(r)\right\rangle
\end{equation}

\subsection{Example 1: Kepler-Coulomb with a 1/$r^3$ perturbation}

This can occur, for example, in an atom with a spin-orbit magnetic
interaction, or in a gravitational system with a Schwarzschild
general relativistic correction \cite{Goldstein}.

The energy of the system is
\begin{equation}
\label{virialkc} E=\langle -kr^{-1}\rangle +\frac 12\langle
kr^{-1}\rangle \end{equation}

If the perturbation is $\Delta V(r)=\lambda /r^3$, the perturbed
energy is
\begin{eqnarray}
\label{perkep}
 E^{\prime }&=&-\frac k2\langle r^{-1}\rangle
+\lambda \langle r^{-3}\rangle
\nonumber \\
  &=&-\frac k{2a}+\frac {\lambda }{b^3}
\end{eqnarray}

which results in a precession of the ellipse.

\subsection{Example 2: Anharmonic oscillator with an $r^4$ perturbation}

The energy of the system is
\begin{equation}
E=\langle \frac 12kr^2\rangle +\langle \frac 12kr^2\rangle
\end{equation}

If the perturbation is $\Delta V(r)=\lambda r^4$, the perturbed
energy is
\begin{eqnarray}
\label{persho}
 E^{\prime }&=&k\langle r^2\rangle +\lambda \langle
r^4\rangle
\nonumber \\
 &=&\frac k2\left( a^2+b^2\right) +\frac {\lambda }4\left[ 3\left( a^2+b^2\right)^2 -2a^2b^2\right]
\end{eqnarray}
which also results in a precession of the ellipse.

\section{The semiclassical EBK quantization}

The semiclassical Einstein-Brillouin-Keller quantization is given
by
\begin{equation}
(n_i+\frac {\mu }4)=\oint  dq_i~p_i
\end{equation}
where $\mu $ the Maslov index, which is the number of turning
points.  This formalism was applied for spherical symmetric
potentials in an earlier paper \cite{EBK}. The angular phase
integrals yield a value for the angular momentum
\begin{equation}
L=(\ell + 1/2)\hbar ~.
\end{equation}
The square of this result
\begin{equation}
L^2 =\left[ \ell (\ell + 1)+ 1/4\right] \hbar ^2
\end{equation}
agrees with the quantum mechanical result in the correspondence
limit.

Our earlier calculations \cite{EBK} for the radial phase integral
permit the specification of the semimajor and semiminor axes of
the ellipses.

\subsection{Kepler-Coulomb}

For the Coulombic atomic problem ($k=Z$e$^2/4\pi \epsilon _0$ for
a hydrogenlike atom),

\begin{eqnarray}
a&=&\frac {\hbar ^2}{mk}(n_r+\ell +1)^2
\nonumber \\
b&=&\frac {\hbar ^2}{mk}(n_r+\ell +1)(l+\frac 12) ~.
\end{eqnarray}

The radial quantum number $n_r$ is displayed here so that the two
potentials can be compared under conditions whereby $n_r$ and
$\ell $ have the same range of values 0, 1, 2, $\dots $ The
expression is usually written in terms of the principal quantum
number $n\equiv n_r+\ell +1$.

With this quantization the perturbed energy of Eq. \ref{perkep}
becomes
\begin{equation}
\label{perekpk} E^\prime=-\frac {k^2m}{2\hbar ^2}\left[ \frac
1{n^2}-\frac {2\lambda }k \frac 1{n^3(l+\frac 12)^2}\right]
\end{equation}
which agrees with the quantum mechanical result with the
correspondence $(\ell +\frac 12)^3 \rightarrow \ell (\ell +\frac
12)(\ell +1)$~.

\subsection{Isotropic harmonic oscillator}

In this case the quantization yields value for the semiaxes
(denoting $\omega \equiv \sqrt {k/m}$)
\begin{eqnarray}
\frac {a^2+b^2}2&=&\frac \hbar {m\omega }\left( 2n_r+\ell +\frac
32\right)
\nonumber \\
ab&=&\frac \hbar {m\omega }\left( \ell +\frac 12\right) ~.
\end{eqnarray}

Here again the radial quantum number $n_r$ is displayed for
comparison with $n_r$ and $\ell $ having the same range of values
0, 1, 2, $\dots $ The expression is usually written in terms of
the quantum number $n\equiv 2n_r+\ell $~.

With this quantization the perturbed energy of Eq. \ref{persho}
becomes
\begin{equation}
\label{perekph} E^\prime=\hbar \omega \left( n+\frac
32\right)+\frac {\lambda \hbar ^2}{2m^2\omega
^2}\left[3\left(n+\frac 32\right)^2-\left(\ell +\frac
12\right)^2\right]
\end{equation}
which agrees with the quantum mechanical result \cite{Ray} with
the correspondence $(\ell +\frac 12)^2 \rightarrow (\ell -\frac
12)(\ell +\frac 32)$~.

\section{Conclusion}

This formulation in terms of the classical position probability
density provides a mathematically simple exposition of the
difference in frameworks between classical and quantum mechanical
physics.  Although this one exercise does not provide a
comprehensive alternative to the standard presentation, it can
clearly illustrate at the introductory level the limitations of
the Newtonian approach.

\appendix

\section{Legendre functions of half-odd-integer order}

Legendre functions of half-odd-integer order can be evaluated
using the hypergeometric series
\begin{equation}
 \label{app1}
 P_{-\nu-1}(z)\, =\, P_\nu(z)\, =\, \left( \frac{1+z}{2}
 \right)^\nu F(-\nu,-\nu; \, 1 \, ; \, \frac{z-1}{z+1}) \, .
\end{equation}
Thus \begin{equation}
 \label{app2}
 \langle r^k\rangle\ =\, \left( \frac{a+b}{2} \right)^k \, F \left(
 -\frac k2, -\frac k2 \, ; \, 1 \, ; \, \left( \frac{a-b}{a+b}
 \right)^2 \right) ~.
\end{equation}
For the case shown in the figures, $b=a/2$, this gives for the
first moment,
\begin{equation}
 \label{app3}
 \langle r \rangle\ =\, \frac{3a}{4}\, F \left( -\frac 12, -\frac
 12 \, ; \, 1 \, ; \, \frac 19\, \right)\, =\, 0.77098\, a ~.
\end{equation}
In the limit $b\rightarrow 0$ we can use the fact that
$F(a,a,;c;1) = \Gamma(c)\Gamma(c-2a) / \Gamma(c-a)^2$ to write the
moments ($k\ge 0$) for a linear oscillator in one dimension:
\begin{equation}
 \label{app4}
 \langle r^k \rangle_{1D}\ =\, \frac{k!\, a^k}{2^k
 \Gamma(1+k/2)^2} ~.
\end{equation}
These results check against the elementary results, for example
\begin{eqnarray}
 \label{app5}
 \langle r \rangle_{1D}&=& 2a/\pi
 \nonumber \\
 \langle r^2\rangle_{1D} &=&  a^2/2
 \nonumber \\
 \langle r^4\rangle_{1D} &=& 3a^4/8 ~.
\end{eqnarray}

\newpage
{\bf FIGURE CAPTIONS}

Figure 1. Comparison of the elliptic orbits with $a=1$ unit and
$b=0.5$ for the two examples.

Figure 2. Classical position probability distributions for the two
elliptic orbits shown in Fig.~1.


\newpage

\begin{thebibliography}{99}

\bibitem{Cambridge}
Lorenzo J.~Curtis, {\it Atomic Structure and Lifetimes: A
Conceptual Approach} (Cambridge, UK, 2003).


\bibitem{Wilczek}
Frank Wilczek,``Whence the force $F=ma$? I: Culture shock,''
Physics Today {\bf 57} (10), 11-12 (2004).

\bibitem{Taylor}
Edwin F.~Taylor, `` A call to action,'' Am.~J.~Phys. {\bf 71} (5),
423-425 (2003).

\bibitem{Martin}
Lorenzo J.~Curtis, Roger R.~Haar, and Martin Kummer, ``An
expectation value formulation of the perturbed Kepler Problem,''
Am.~J. Phys. {\bf 55} (7), 627-631 (1987).

\bibitem{Gradsteyn}
I.~S.~Gradshteyn and I.~M.~Ryzhik, {\it Tables of Integrals,
Series and Products}, (Academic, New York, 1965), formulas 3.661-3
and 3.661-4.

\bibitem{Sivard}
Jean Sivardi\`ere, ``Laplace vectors for the harmonic
oscillator,'' Am. J. Phys. {\bf 57} (6), 524-525 (1989).

\bibitem{Goldstein}
Herbert Goldstein, {\it Classical Mechanics} (Addison-Wesley,
Reading MA, 1980) 2nd ed., p.~511

\bibitem{EBK}
Lorenzo J.~Curtis and David G.~Ellis, ``Use of the
Einstein-Brillouin-Keller action quantization,'' Am.~J.~Phys. {\bf
72} (9), 1521-1523 (2004).

\bibitem{Ray}
Aparna Ray, Kalyaneswari Mahata, and Pritam P.~Ray, ``Moments of
probability distribution, wavefunction, and their derivatives at
the origin of $N$-dimensional central potentials,'' Am. J. Phys.
{\bf 56} (5), 462-464 (1988).

\end{thebibliography}
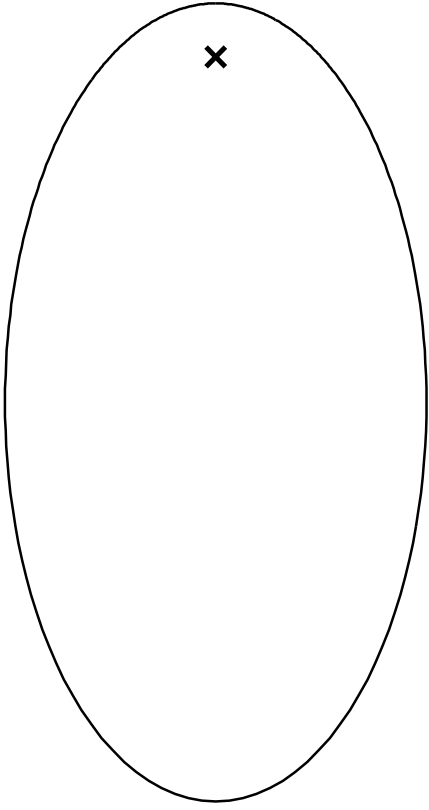
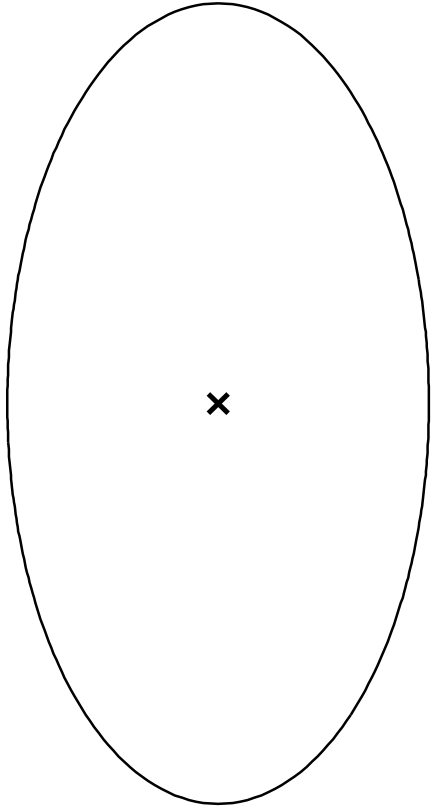
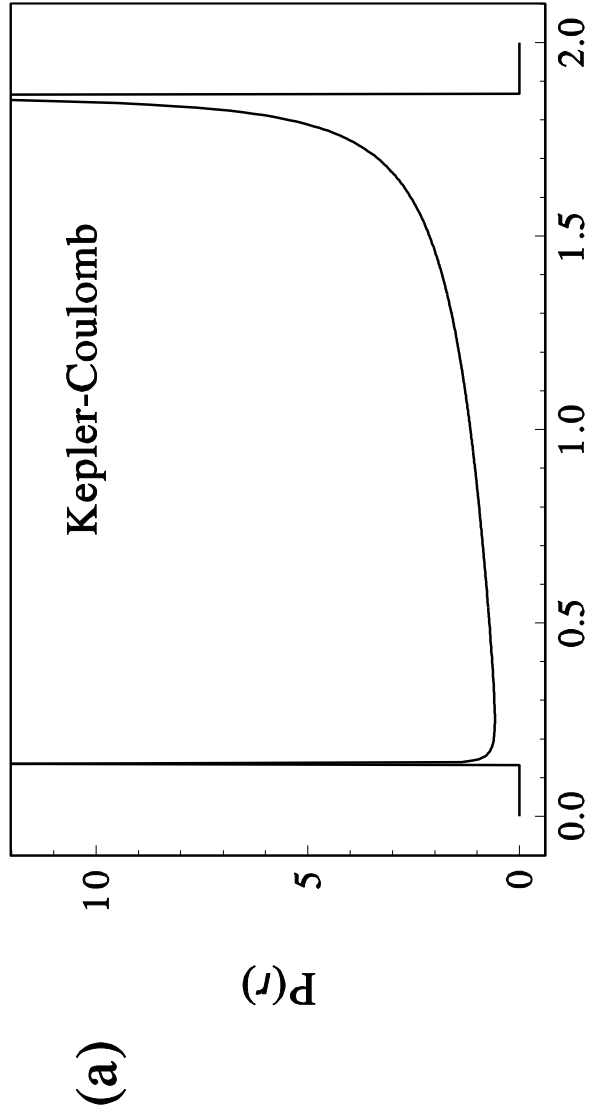
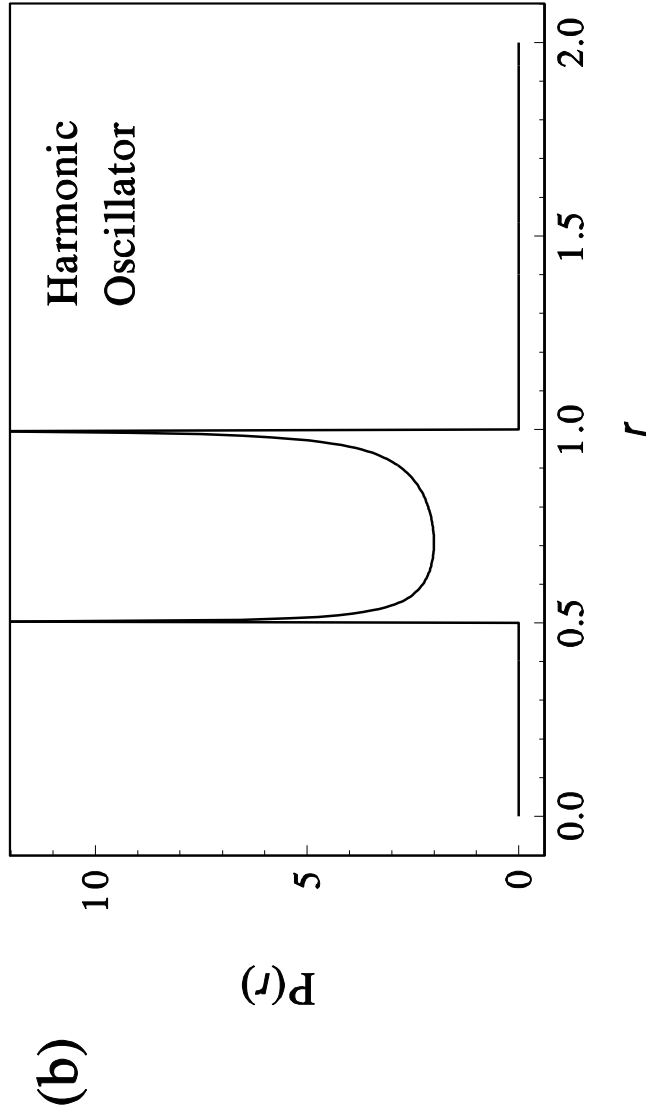
\end{document}